\documentclass[journal,comsoc,twoside]{IEEEtran}
\PassOptionsToPackage{draft}{hyperref}
\PassOptionsToPackage{obeyspaces}{url}

\usepackage{float, enumerate, xcolor, algorithm, algorithmic, tabto, balance, amsmath, graphicx, epstopdf, hyperref, multirow, dblfloatfix, textcomp, listings, eurosym, lstautogobble, multirow, tikz} 
\usepackage{amsfonts}

\usepackage[inline]{enumitem}

\usepackage{amssymb}
\usepackage{booktabs}
\usepackage{url}
\usepackage{amssymb}
\usepackage[numbers,sort&compress]{natbib}
\usepackage[flushleft]{threeparttable}
\usepackage{subfigure}
\usetikzlibrary{arrows,backgrounds,positioning}
\usepackage{chngcntr}
\usepackage[overload]{empheq}
\usepackage{soul}

\ifCLASSINFOpdf
\else
\fi

\newcommand{\ie}{{i.e.,~}}
\newcommand{\eg}{{e.g.,~}}

\newcommand{\vs}{{vs.~}}

\newcommand{\ankit}[1]{\textcolor{black}{#1}}
\newcommand{\michele}[1]{\textcolor{black}{#1}}
\newcommand{\rankit}[1]{\textcolor{black}{#1}}
\newcommand{\rmichele}[1]{\textcolor{black}{#1}}
\newcommand{\rrankit}[1]{\textcolor{black}{#1}}
\newcommand{\rrmichele}[1]{\textcolor{black}{#1}}

\usepackage{array}
\newcolumntype{L}{>{\centering\arraybackslash}m{.8\columnwidth}|}

\usepackage{url}

\usepackage{pifont}

\usetikzlibrary{arrows,automata}

\graphicspath{ {./images/} }

\begin{document}
\title{Blockchain Trilemma Solver Algorand has Dilemma over Undecidable Messages}
	\author{Mauro~Conti,
		~Ankit~Gangwal*,
		~Michele~Todero$\dag$
        \IEEEcompsocitemizethanks{
            \IEEEcompsocthanksitem All authors are with the Department of Mathematics, University of Padua, 35121, Italy.
            \IEEEcompsocthanksitem \textit{* Corresponding author: Ankit Gangwal (ankit.gangwal@phd.unipd.it)}
            \IEEEcompsocthanksitem \textit{$\dag$ This work is an outcome of Michele Todero's M.Sc. thesis.}
        						}
		%
		}
\markboth{
}{Conti \MakeLowercase{\textit{et al.}}: Blockchain Trilemma Solver Algorand has Dilemma over Undecidable Messages}


\maketitle

\begin{abstract}
\ankit{A variety of solutions, \eg Proof-of-Work (PoW), Proof-of-Stake (PoS), Proof-of-Burn (PoB), and Proof-of-Elapsed-Time (PoET), have been proposed to make consensus mechanism used by the blockchain technology more democratic, efficient, and scalable. However, these solutions have a number of limitations, \eg PoW approach requires a huge amount of computational power, scales poorly, and wastes a lot of electrical energy.}
\par 
\ankit{Recently, an ingenious protocol called Algorand has been proposed to overcome these limitations.} Algorand uses an innovative process - called cryptographic sortition - to securely and unpredictably elect a set of voters from the network periodically. These voters are responsible for reaching consensus through a Byzantine Agreement (BA) protocol on one block per time, guaranteeing an overwhelming probability of linearity of the blockchain.
\par
\ankit{In this paper, we present a security analysis of Algorand. To the best of our knowledge, it is the first security analysis as well as the first formal study on Algorand. We designed} an attack scenario in which a group of malicious users tries to break the protocol\ankit{, or at least limiting it to a reduced partition of network users, by} exploiting a possible security flaw in the messages validation process of the BA. \ankit{Since, the source code or an official simulator for Algorand was not} available at the time of our study, we created a simulator (which is available on request) to implement the protocol and \ankit{assess the feasibility of our attack scenario.} \rankit{Our attack requires the attacker to have a trivial capability of establishing multiple connections with targeted nodes and costs practically nothing to the attacker. Our results show that it is possible to slow down the message validation process on honest nodes, which eventually forces them to choose default values on the consensus; leaving the targeted nodes behind in the chain as compared to the non-attacked nodes.} \ankit{Even though our results are subject to real-implementation assumption, the core concept of our attack remains valid.}
\end{abstract}


\IEEEpeerreviewmaketitle

\section{Introduction}
\IEEEPARstart{W}{ith} \ankit{the introduction of Bitcoin~\cite{nakamoto2008bitcoin} in 2008, blockchain technology has been adopted by almost every scenario in today's digital world with an aim to keep temper-proof records~\cite{conti2018economic}. Originally, the concept of distributed consensus - upon which the blockchain technology is founded - dates back to the~'90s~\cite{haber1990time, szabo2005bitgold, Wei1998money, finney2004rpow}. But, after the success of Bitcoin, a multitude of other areas have been} transformed by the idea of a transparent and distributed public ledger, such as elections, contract management, insurance, charity, and ownership transfer through inheritance and will~\cite{blockchain_app}.
\par
The Bitcoin system combines together different well-known concepts, such as digital signatures, hashing functions, PoW, and Merkle trees. However, \ankit{with an exponential growth in the number of users and Bitcoin-miners,} the Bitcoin network observed collateral effects of its consensus protocol. \ankit{In particular, it suffers from} scalability~\cite{croman2016scaling}, hashing power centralization, mining strategies exposures~\cite{rosenfeld2011analysis, eyal2015miner, luu2015power}, and energy- and computational-deficiency~\cite{malone2014energy}. \ankit{To this end, a} variety of solutions have been proposed over time. Some of these solutions \ankit{aim to improve the existing version of PoW (\eg Litecoin\footnote{Litecoin adopted its \textit{scrypt} PoW function's design to increase the decentralization of the system and further optimize the mining process~\cite{percival2009stronger}.}, Primecoin\footnote{Primecoin exploits the computational power used for the mining activity to contribute to mathematical research topics.}~\cite{king2013primecoin}}) while other proposals used completely different concepts, such as PoS, PoB~\cite{wiki_pob, slimcoin}, and PoET~\cite{intel}. \ankit{As a representative example, PoS systems (both stake modification-based~\cite{king2012ppcoin, cloakcoin2014, ren2014proof, pikepost} and actual balance-based~\cite{nxt, vasin2014blackcoin}) have serious concerns from the security point-of-view~\cite{li2017securing}; such issues have been partially solved by providing formally-proved security guarantees~\cite{kiayias2017ouroboros, bentov2016snow} or by discouraging users to act maliciously through punitive procedures~\cite{buterin2017casper, buterin2014slasher}. For the sake of completeness, some other works~\cite{sompolinsky2016spectre, popov2016iota} tried to achieve better scalability and throughput by implementing a block graph structure based on Directed Acyclic Graph (DAG), albeit, these solutions have their own demerits.}
\par
\ankit{On the other side, Algorand~\cite{micali2016algorand_tec, micali2017algorand_pap} - an innovative protocol for the blockchain technology -} \rankit{aims to solve the ``blockchain trilemma'' of decentralization, scalability, and security.} It relies on PoS integrated in a BA layout to achieve optimal resilience against faulty nodes. It guarantees an overwhelming probability of the blockchain's linearity and a block rate of nearly a minute. It is comparable to the approaches presented in~\cite{kiayias2017ouroboros, bentov2016snow} \ankit{and has the potential to shape the future of the blockchain technology.}
\par
\textit{Motivation:}
Given the promising properties of the Algorand protocol in terms of decentralization and scalability, the security aspects of \ankit{this consensus algorithm} are crucial. To the best of our knowledge, our work is the first analysis of Algorand's security. We aim to at least start a discussion about a possible security flaw on the message validation process, which could possibly expose honest nodes' bandwidth and memory resources to Distributed Denial-of-Service \ankit{(DDoS)} attacks. 
\par
\textit{Contribution:} 
The major contributions of our work are as follows:
\begin{enumerate}
\item \rankit{We demonstrate a practically feasible attack on the Algorand protocol. The attack has the potential to target an arbitrarily sized group of honest users and slow down the consensus process; leaving the targeted nodes behind in the chain as compared to the non-attacked nodes.}
\item We implemented the protocol in our simulator created in Java and evaluated the feasibility of our attack. The simulator is available on request.
\end{enumerate}
\textit{Organization:} 
The rest of the paper is organized as follows: Section~\ref{section:algorand} gives a thorough description of the Algorand protocol. We elucidate our attack scenario in Section~\ref{section:attack}. Section~\ref{section:evaluation} presents the implementation details of our simulator, evaluation settings, and our results. Section~\ref{section:analysis} discusses the feasibility of our attack and the possible countermeasures. Finally, Section~\ref{section:conclusion} concludes the paper by listing potential future works.

\section{Algorand}
\label{section:algorand}
\michele{Since the main focus of our work is on Algorand, we first elucidate the Algorand protocol. We introduce the main limitations of current blockchain technologies in Section~\ref{section:limits}. We present the distinct features of Algorand in Section~\ref{section:properties}. Section~\ref{section:assumptions} discusses the assumptions specified by the Algorand protocol with respect to the adversary model. Section~\ref{section:network} gives an overview of the network topology, communication protocol, and the definition of messages. Then, consensus mechanism of Algorand is described in Section~\ref{section:consensus}. Finally, the technical details of ``cryptographic sortition'' - a critical component of the consensus process - are discussed in Section~\ref{section:sortition}.}

\subsection{The limitations of current Blockchain} 
\label{section:limits}
\ankit{The main motivation for the development of Algorand lies in the underlying assumption and technical problems that
are essentially shared by most of the PoW-based blockchains~\cite{micali2016algorand_tec}. Those limitations include:
\begin{enumerate}
	\item $\lbrack$Assumption$\rbrack$ Majority of nodes contributing to the computational power of the network will be honest.
	\item $\lbrack$Technical Problem$\rbrack$ Wastage of computational power as well as electrical energy.
	\item $\lbrack$Technical Problem$\rbrack$ Concentration of power, \ie the total computing power for block generation lies within just few mining pools.
	\item $\lbrack$Technical Problem$\rbrack$ Ambiguity in the consensus reached by different network nodes on the confirmed transactions, which leads to fork of blockchain.
\end{enumerate}
}

\subsection{Salient features of Algorand}
\label{section:properties}
The protocol has distinct and promising features:
\begin{enumerate}
	\item It is designed to be fully decentralized and democratic. \michele{Moreover, there is no distinction between the role of different groups of users, \eg miners \vs ``normal'' users.}
	\item Each user runs the same functions with negligible hardware requirements (\ie with negligible computational effort) \ankit{as the concept of miner/mining does not exist in Algorand.}
	\item \ankit{It is scalable with the number of nodes as well as in terms of confirmed transactions throughput. The authors~\cite{micali2017algorand_pap} have shown that the throughput of Algorand is fifty times that of Bitcoin.}
	\item The blockchain does not fork with overwhelming probability ($P_{fork}=10^{-12}$).
\end{enumerate}

\subsection{Protocol assumptions}
\label{section:assumptions}
\ankit{To guarantee security of the blockchain, Algorand relies on the conventional assumption of PoS systems, \ie Honest Majority of Money~(HMM). In particular, Algorand assumes a continuum in the ownership of currency belonging to honest users, }\michele{\ie at any round $r$ at least a fraction $h$ (greater than two-third) of the money held by honest accounts on round $r-k$ \ankit{must still belong to the} honest users, for some integers $h$ and $k$ that are \ankit{defined by} the protocol.} 
Together with HMM, Algorand requires a partial synchronization assumption on the honest nodes' clocks. It states that the honest nodes' clocks are not required to be strictly synchronized, but they are assumed to have the same speed with a bounded tolerance. 
\par
\ankit{Furthermore}, Algorand was built to face a very strong adversary. The adversary is capable to perfectly synchronize all malicious nodes \ankit{that} he\footnote{In this paper, we have mentioned different entities as a masculine entity without any gender-bias.} controls and can corrupt honest users indiscriminately and instantly. \ankit{However, the attacker's capabilities are bounded by the following three main facts}: (1)~he can corrupt honest users until the HMM property is maintained; (2)~he cannot forge secret keys of the honest users that he have not corrupted; and (3)~\ankit{he cannot prevent the honest recipient to receive messages.}

\subsection{Network communication}
\label{section:network}
The Algorand network is a Bitcoin-like peer-to-peer network. Each user is identified by a public/private key pair, and each key pair corresponds to a node in the topology. \ankit{For simplicity, we will use the terms ``user'' and ``node'' interchangeably}. Each node establishes and maintains a different TCP connection with each of his peers. \ankit{The current specifications of Algorand do not specify if a node has to choose his peers based on their stake, which indeed exposes the network to a Sybil attack\footnote{\ankit{In a sybil attack, the attacker creates a large number of nodes at zero cost (\ie with no stake) to increase the probability of connecting with his targets.}}.}
\par
Nodes broadcast messages through the network using a standard gossip protocol - each node gossips his message to his peers, who in turn relays it to their neighbors. A message has to be validated before it can be relayed, and it is never sent twice to the same user. 
The protocol defines four types of messages: (1)~transactions; (2)~voting messages; (3)~block proposals; and \rmichele{(4)~credential messages}. A transaction (t) from user $i$ to user $j$ is defined as:
\begin{equation}
t=sig_i(pk_i, pk_j, a, I, H(SI)),
\end{equation}
where $pk_u$ is the public key of the user $u$, $a$ stands for the amount of Algorand units to transfer from $pk_i$ to $pk_j$, $I$ is an optional string to describe the transfer, $H(SI)$ is a 256-bit hash produced by hashing 
a secret string $SI$ (not specified in the body of the message), and  $sig_u(x)$ \ankit{denotes digitally signed data~$x$ using the private key of user $u$.}
 \par
A voting message \ankit{(vm)} from a user $i$ in round $r$ and step $s$ is defined as:
\begin{equation}
\label{definition:voting_message}
vm_i^{r,s}=(sig(v_i^{r,s}), \sigma_i^{r,s}),
\end{equation}
where $\sigma_i^{r,s}$ represents the sortition proof of the \ankit{user $i$ in $s^{th}$~step of $r^{th}$~round} (further details in Section~\ref{section:sortition}), $v_i^{r,s}$ defines the vote of user $i$ in round $r$ and step $s$. 
\par
\michele{A block proposal \ankit{(bp)} for round $r$ from user $i$ is defined as:
\begin{equation}
\label{definition:block_proposal}
bp^r_i=(B^r_i, sig_i(H(B^r_i)), \sigma^{r,1}_i),
\end{equation} 
where $B^r_i$ is defined as 
\begin{equation}
B^r_i=
 \begin{cases}
   (r, PAY^r, sig_i(Q^{r-1}), H(B^{r-1})), & \quad \text{if } B^r_i\neq E^r;\\
    (r,Q^{r-1},H(B^{r-1})),  & \quad \text{otherwise;}
  \end{cases}
\end{equation}
where $PAY^r$ is the set of transactions $t$, while $Q^r$ is called \textit{seed} and defined as 
\begin{equation}
\label{definition:seed}
Q^r =
  \begin{cases}
    H(sig_{l^r}(Q^{r-1},r)), & \quad \text{if } B^r\neq E^r;\\
    H(Q^{r-1},r),  & \quad \text{otherwise;}
  \end{cases}
\end{equation}
where $l^r$ is the user who created the consensus block $B^r$ for round $r$. The seed is a parameter needed in the sortition process (more details in Section~\ref{section:sortition}). $E^r$ is the default block for round $r$, and it is defined as
\begin{equation}
E^r=(r,Q^{r-1},H(B^{r-1})).
\end{equation}
}
\par
\rmichele{Finally, the credential message (cm) \rankit{from a user $i$ for a round $r$} is defined as:
\begin{equation}
\label{definition:credentials}
cm^r_i=(\sigma^{r,1}_i).
\end{equation}
It consists of the sortition proof of the block proposer (\rankit{in this case, user $i$}). Each sortition proof is associated with a priority value in a deterministic way \michele{(further details in Section~\ref{section:sortition})}. Credential messages are far smaller in size than the block proposals. Thus, they propagate faster through the network. Credential messages are gossiped by the block proposers at the beginning of a round along with their block proposals. Since the block proposals and the credential messages from the same block proposer contain the same sortition proof, the peer nodes can leverage the priority values from the credential messages to not relay block proposals with low priorities - which helps in preventing congestion in the network.
}

\subsection{Consensus algorithm}
\label{section:consensus}
Algorand's consensus algorithm consists of a synchronous protocol that combines the concepts of PoS systems with a Byzantine fault tolerance agreement. The protocol virtually schedules the time into rounds. At each round, all the network nodes attempt to reach consensus on a new block of transactions. Each round is composed of the following actions:
\begin{itemize}

\item Each node in the network must first check its role to determine whether he has to propose a block for a given round. To do so, the nodes use cryptographic sortition, which requires them to run a trivial (a single hash) computation challenge. In case a node has to propose a block, he collects all the pending transactions inside a block proposal and gossips it together with the sortition proof. The block's size is limited by a fixed protocol parameter.

\item \michele{Each node waits for incoming block proposals from the previous step for a predefined duration of time. Among all the valid blocks he receives, he selects the one with the highest priority sortition proof. Then, he computes an hash using this block, sets this hash as the input for the BA of the next steps.}

\item All nodes in the network try to reach consensus on one block through a BA protocol. The BA has two key phases. Both the phases are composed by subsequent steps. At each step, a distinct group of users (called committee members) is elected in an unpredictable way through cryptographic sortition. These committee members spread their voting messages based on votes received from the previous step. The committee members attach a sortition proof that proves their legitimacy while spreading their voting messages. The two phases of BA are as follows:
\begin{enumerate}
    \item The first phase is called the \textit{reduction} phase that comprises two steps. \ankit{In the first step, each committee member votes for the hash of the blocks proposed for consideration. In the second step, committee members vote for the hash that received votes over a certain threshold (defined by protocol). In case none of the hash/block receives enough votes, committee members vote for the hash of the default empty block.
    } 
    \item The next phase consists of another binary BA to ensure that each node agrees on the same consensus; the consensus can conclude on the output of the previous phase \rankit{or} on the default value in the absence of a majority.
\end{enumerate}
\end{itemize}

\subsection{Cryptographic sortition}
\label{section:sortition}
\michele{Cryptographic sortition is used by each network node at the beginning of every step of each phase to determine whether he has a role in that particular step. \ankit{It is a simple and lightweight process that involves computing a single hash and a digital signature}. The sortition algorithm is implemented using a Verifiable Random Function (VRF\footnote{\ankit{VRF functions allow users to produce verifiable computations. The users produce a proof using their private keys, which can be verified by other users via the producer's public key.}}
)~\cite{micali1999verifiable}}. In particular, the hash for a certain step $s$ and round $r$ is computed by the user~$i$ as
\begin{equation}
y=H(sig_i(r,s,Q^{r-1})),
\end{equation}
\michele{where $H$ is a hashing function, $y$ a 256-bit long hash, and $Q^{r-1}$ is the seed quantity defined by Eq.~\ref{definition:seed}. $sig_i(r,s,Q^{r-1})$ is called the sortition proof and is attached to voting messages or block proposals to prove the legitimacy of the corresponding voters or block proposers. The hash is used by the nodes to verify a certain sortition condition. The possibility that the condition is verified is directly proportional to the stake of the node and depends on a constant parameter \michele{$\pi_{role}$} fixed in the protocol, which guarantees that on average a fixed number of nodes are elected at each step for a certain role. For this reason, the protocol allows users to be elected more than once in the same step. A user $i$ can possibly be elected at each step $w_i$ times, where $w_i$ is the number of user's units of currency. Formally, the interval $[0,1)$ is partitioned into $w_i$ intervals \michele{$I$, where the $j$-th interval $I_j$ is defined as:}}
\begin{equation}
I_j=\Big[\displaystyle\sum_{x=0}^j B(x; w_i; p), \displaystyle\sum_{x=0}^{j + 1} B(x; w_i; p)\Big), j \in (0,1,...w_i)
\end{equation}
\michele{where $B(x; w_i; p)$ is the binomial probability that user~$i$ is elected exactly $x$ times on $w_i$ chances, each with probability $p=\pi_{role}/{W}$. Here, $\pi_{role}$ estimates the expected number of sorted users for that role and $W$ is the total amount of currency in the network. The number of votes are found by applying the binary point operator to the hash (\ie $y/2^{y.length}$) and searching the interval $I_j$ with which the value is associated. If the value lies on the $j$-th interval, then $i$ can vote $j$ times.
There are at least three interesting observations about this procedure: 
\begin{enumerate}
\item Since on average a fixed number of nodes are elected at each step, the traffic of voting messages in the network does not increase with the number of participants.
\item The procedure to decide whether a user~$i$ is sorted for a certain step in a round can be executed only by the user~$i$ as it involves a signature process.
\item Since $Q^{r-1}$ can be computed by a node only when that node reaches consensus on round $r-1$, the result of the sortition of user~$i$ in a certain step of round $r$ is unpredictable and unknown even to the user~$i$. This avoids the possibility of collusion between voters of one or more steps to manipulate the consensus process on a certain round.
\end{enumerate}}

\section{Our attack}
\label{section:attack}
\michele{We explain our attack in this section, starting from attack preliminaries in Section~\ref{section:attack_assumption} followed by a brief introduction to flooding attacks in Section~\ref{section:attack_scenario}, and the description of our attack in Section~\ref{section:attack_undecidable}.}

\subsection{Attack preliminaries}
\label{section:attack_assumption}
In our adversary model, we assume that all the malicious nodes are coordinated by a single/unique entity. \michele{This translates into the following two properties of the malicious nodes:
\begin{enumerate}
\item Each malicious node can sign messages using any other malicious node's private key.
\item The malicious nodes participate in the protocol either passively or actively. These malicious nodes coordinate and use the messages they receive from their honest peers to be aware of the protocol's execution status.
\end{enumerate}
We assume that all the malicious nodes and their public/private key pairs were created by the adversary sufficiently in advance, \ie before the attack takes place. This enables all of them to become the voters or block proposers in the protocol}\footnote{This assumption is necessary because Algorand increases the unpredictability of the outcome of sortition process by limiting the set of keys that can participate in the process. The eligible keys must be created at least $k$ rounds before.}. As mentioned in Section~\ref{section:network}, we can also safely assume that the network is vulnerable to the Sybil attack, meaning that the honest nodes choose their peers randomly without relying on their stake. In this way, the malicious nodes have the same probability as of the other nodes to connect to honest peers. \michele{ On the other side, honest peers can control/limit the number of incoming connections, denoted by \textit{max-connections}.}

\subsection{A typical flooding attack}
\label{section:attack_scenario}
\ankit{In a typical network scenario, a flooding attack targets honest nodes' bandwidth and memory resources by sending a huge number of message to him.} Typically, the attack is performed at the application layer and messages are sent over previously established connections. Generally, a flooding attack sends a huge number of fake or invalid messages towards the target nodes to slow down their message reception and message processing. Crafting such attacks from a single nodes is not very effective because nodes in Bitcoin-like decentralized networks - where no level of trust is assumed between participants - often rely on ``misbehaviour'' scores to label and identify possible malicious users. As soon as an honest node decides that one of his peers is malicious, he can stop processing messages from that peer, drop the connection, and blacklist him to prevent future connection requests from that peer for a certain time. 
\ankit{Hence, the attacker needs a large number of malicious nodes connected to the target. So he can send some messages from each connection in a coordinated manner and create the same impact without being detected (or at least fully detected).}

\subsection{Magnifying attack's impact using undecidable messages}
\label{section:attack_undecidable}
\ankit{Our attack aims to increase the impact of the flooding attacks by exploiting the message validation process of Algorand. Due its design, our attack also enables an adversary to have a longer connection time before getting blacklisted by the target node.}
\michele{In Algorand, each message that is a part of the consensus algorithm (\ie voting messages (Eq.~\ref{definition:voting_message}) and block proposals (Eq.~\ref{definition:block_proposal})) is subject to two distinct checks during the validation process:
\begin{enumerate}
\item \textit{Stateless checks}: A set of checks on a message/packet that a validator node can perform to know its current status (\eg packet's structure, packet's authentication, check for duplicated messages). 
\item \textit{Sortition proof check}: The sortition proof must be checked to verify that the sender/author of the message had the right to gossip it.
\end{enumerate}
As described Section~\ref{section:sortition}, the sortition proof is a byte-string that represents the signature of a user $i$ in the form $sig_i(r,s,Q^{r-1})$}, the validator node should hash this string to verify whether the given output satisfies the sortition property. However, before blindly hashing the signature, the protocol requires him to verify it first. Thus, he needs all the parameters - that were originally used to create it - including the seed quantity $Q^{r-1}$. This is a stateful check because the validator should have already reached consensus on the round $r-1$ to independently compute $Q^{r-1}$. If that is not the case, then the message cannot be fully validated at the current status of the validator node. In such a situation, the only option that the validator node has is to store the message and wait until it can be fully validated because messages that are not yet fully validated, cannot be discarded. We refer to these messages as \textit{undecidable messages}.
\par
\michele{At this point, the aim of the attacker is to flood honest targets with as many undecidable messages as possible to saturate their bandwidth and possibly their memory. These two objectives decide the type and number of messages used to construct our attack vector. With respect to the type, we chose the largest possible message, \ie block proposals containing max-sized blocks. Since messages need to pass the stateless checks, each public key can gossip only a single block proposal per round. However, keys are created at zero cost and each malicious node can sign messages with any of the adversary keys. Moreover, each node can include an arbitrary number of block proposals in his payload - one for each key\footnote{As long as the messages are authenticated and sent only once, a single public key is able to create a different undecidable message for every step in each round. It means that it can create a block proposal on the first step and a voting message for every step of BA (until the maximum $m$-th step allowed by the protocol). However, sending one message for every step implies that the given public key is sorted at every step of that round, which is very unlikely. A similar argument can be used by the target node to statistically label malicious users without waiting for their messages to be fully verifiable. Moreover, given a reasonable max-block size (order of MB) and $m$ set to 150 steps~\cite{micali2017algorand_pap}, a single block proposal containing a max-size block would make up the large majority of the total payload weight. Hence, choosing a single block proposal for each key would imply that each user is elected only once in a given round, and cheaters are more difficult to identify a-priori.}.}
\par
\michele{Now, we describe the attack's execution in practice. The attacker connects to his target as one of his peers. Since the target node is assumed to honestly execute the Algorand protocol, as soon as he receives and validates a message for a certain round $r$, he has to relay it to all of his peers including the malicious node. Hence, the attacker just needs to (1)~prepare its payload in advance\footnote{The attacker does not need any information to pre-compute a payload for a certain round $r$. He just needs to sign some random messages declaring them as if they belong to round $r+1$. All additional information will be checked only once the message becomes fully verifiable in round $r+1$, by that time the attacker would have already achieved his goal.} for the attack in round $r$; (2)~wait to receive a block proposal \rmichele{or a credential message~(Eq.~\ref{definition:credentials})} from his target\footnote{Block proposals together with the credential messages are gossiped first through the network at each round.} in round $r$ to be aware of the fact that the target has started the execution of round~$r$; and (3)~send the payload to the target \rankit{from each connected malicious node}.}

\section{Evaluation}
\label{section:evaluation}
\michele{In this section we present the evaluation of our proposed attack. Section~\ref{section:implementation} provides the technical details of our simulator while the evaluation settings and the results are presented in Section~\ref{section:evaluation_settings} and Section~\ref{section:results}, respectively.} 

\subsection{Simulator}
\label{section:implementation}
\ankit{Since, the source code or an official simulator for Algorand was not available at the time of our experiments, we built our own simulator that is completely written in Java. Both the works~\cite{micali2016algorand_tec, micali2017algorand_pap} on Algorand propose slightly different, but overlapping versions of the protocol. The differences are in the terminating conditions to reach the final majority consensus within BA. However, our threat model is independent of the particulars of the chosen version of BA. We referred to the protocol described in the technical paper~\cite{micali2017algorand_pap} for the implementation in our simulator.}
\par
We model an honest node as an entity composed of three main components: \michele{(1)~a network interface to \ankit{send/}receive messages from peers; (2)~a validator process to verify the received messages; and (3)~the process that runs the protocol}. Since our attack focuses on flooding block proposals or voting messages, transactions are never sent/received directly in our simulation. Instead, the throughput (transactions confirmed per second) is simulated by creating block proposals containing a fixed number of \ankit{arbitrary} transactions\michele{\footnote{We assume that in a real implementation of the protocol, messages and transactions are processed by independent processes.}}. 
\par
To reduce the impact of simultaneous/parallel signature verification during simulation, we simulate block validation by using a sleep procedure. Furthermore, nodes are provided with a cache memory not to verify the same transaction twice\footnote{Without a dedicated cache, transactions are validated at least twice. The time when the transaction is gossiped by its original creator and the time when it is received in a block.}. 
\rankit{While block signatures are simulated other checks, \eg packet authentication and sortition proof validation, are actually computed legitimately.} \ankit{We do not assume any latency in the network communication for the sake of presenting the minimum impact of our proposed attack, which would further enhance in the presence of latency.}

\subsection{Evaluation settings}
\label{section:evaluation_settings}
All simulations were done on a virtual machine on OpenStack~\cite{openstack} running Ubuntu 16.04 with 16 Intel Core Processors (Hashwell, no TSX, IBRS) and 64 GB of RAM. In our experiments, we simulated a network of \rrmichele{500} nodes, where each node had a 30 Mbps upload and 30 Mbps download bandwidth. \ankit{The target nodes were connected to all the malicious nodes involved in the attack. \rankit{However, the number of malicious nodes does not represent the \textit{max-connections} parameter controlled by the honest nodes as honest nodes were connected to malicious as well as other honest peers.} For the sake of simplicity, we assumed that once an undecidable message fails in the verification process, then all messages from the sender of that message are discarded without being processed.} \rrankit{Each simulation was run for $\sim$45 minutes, and the attack payload was prepared using different blocks sizes during different experiments.}
\par
To evaluate the effectiveness of our attack, we primarily focus on the number of \ankit{``legitimate''} messages received and validated in due time by the honest nodes \ankit{as well as on the average time taken to complete a round}. \michele{The two evaluation metrics are correlated in a way that if too many messages are not received or \ankit{processed} in due time for a given step, the execution time of the corresponding step increases until a timeout is reached.} We thoroughly evaluated our attack scenario by varying the number of malicious nodes involved in the attack as well as the number of block proposals sent from each malicious node. In particular, the maximum number of tested block proposals (sent from one malicious node at the time of attack) was 70, following the analysis of the cryptographic sortition for block proposers by the authors of the protocol~\cite{micali2017algorand_pap}. \ankit{We set the block proposal receiving time and the step timeout (explained in the next section) to 150 and 60~seconds, respectively.} \rrmichele{Finally, we also evaluated the impact of our attack by varying the size of the attack payload.}

\subsection{Results}
\label{section:results}
\michele{
}
\par 

\begin{figure}[H]
	\centering
	\includegraphics[trim = 2mm 2mm 2mm 2mm, clip, width=.95\columnwidth]{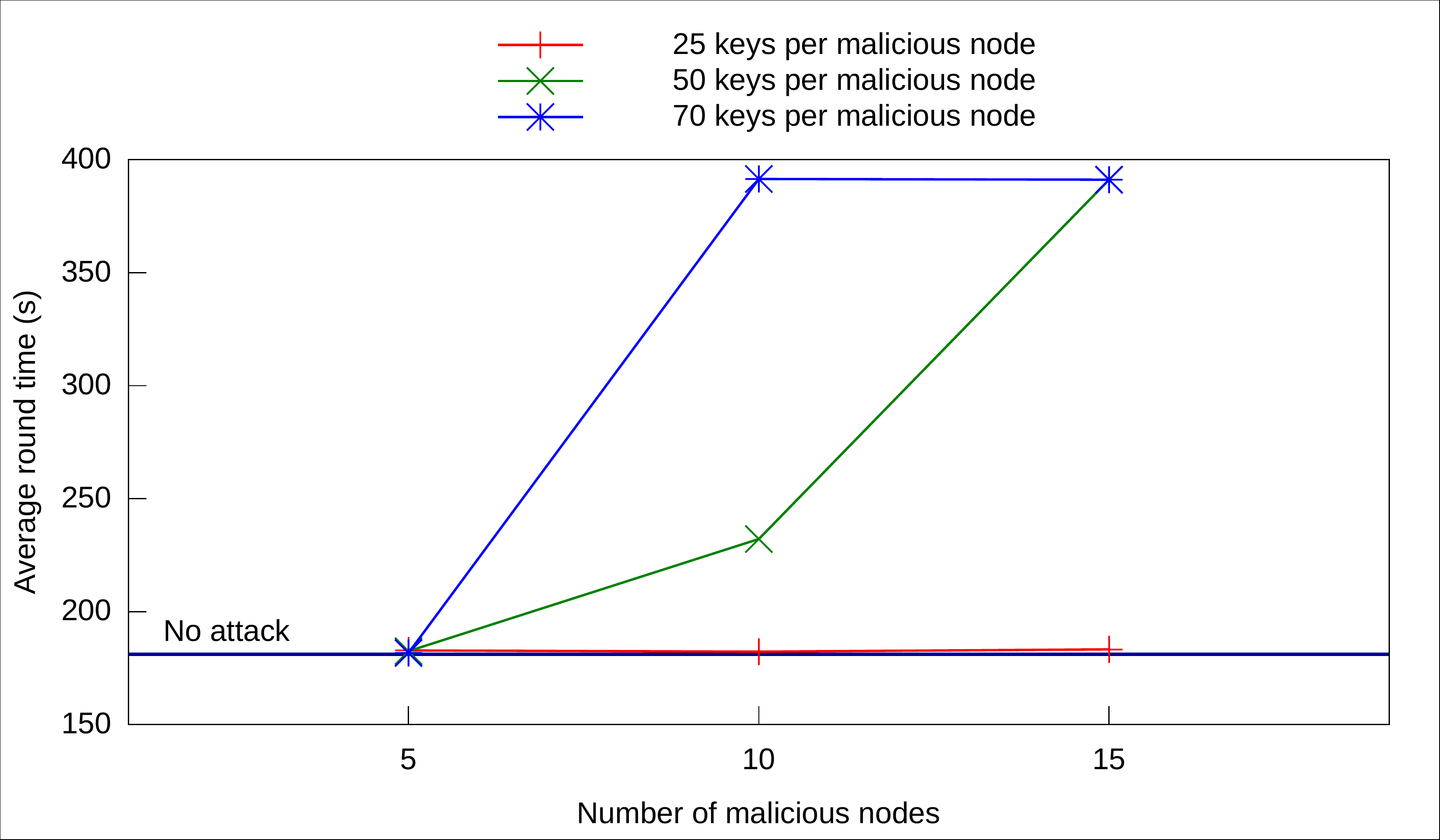}
	\caption{\ankit{Effect of the number of malicious nodes and keys per malicious nodes upon average round time
	}}
	\label{chart:time}
\end{figure}

\ankit{The execution time for a round consists of two parts: block proposal receiving time and step timeout. The block proposal receiving time is constant and is defined as the maximum window of time for the nodes to receive \rankit{the block proposals for a given round}. The step timeout is the maximum amount of time by which each step must complete. Here, each step finishes as soon as enough voting messages are processed or the step timeout occurs.} In ``no attack'' scenario, the largest contribution to round time should come from the block proposal receiving time and the steps should execute before the step timeout occurs. The same is reflected in our results shown in \figurename~\ref{chart:time}. 
Here, in the absence of malicious nodes, 
nearly \rrankit{83}\% of the round time was taken by the fixed block \rankit{proposal} receiving window. \ankit{On the other side, the round time remains nearly the same as ``no attack'' scenario with a lower number of malicious nodes/keys per malicious node. However, the protocol's performance starts to deteriorate with increasing number of malicious nodes/keys per malicious node, starting from the attack payload size of just 500 blocks (10 malicious nodes with 50 malicious keys). \rankit{In facts, the average round time with all the remaining higher attack configurations was over \rrmichele{390}~seconds; which means that these configurations caused at least \rrmichele{four} step timeouts in addition to the block proposal receiving time.} As mentioned earlier, \rankit{our proposed attack does not prevent nodes from reaching consensus. Instead, it aims to significantly increase the execution time of each round, which eventually leads the attacked nodes to reach their consensus on the default value, leaving the targeted nodes behind the non-attacked nodes in the chain.}}
\par
\begin{figure}[H]
	\centering
	\includegraphics[trim = 2mm 2mm 2mm 2mm, clip, width=.95\columnwidth]{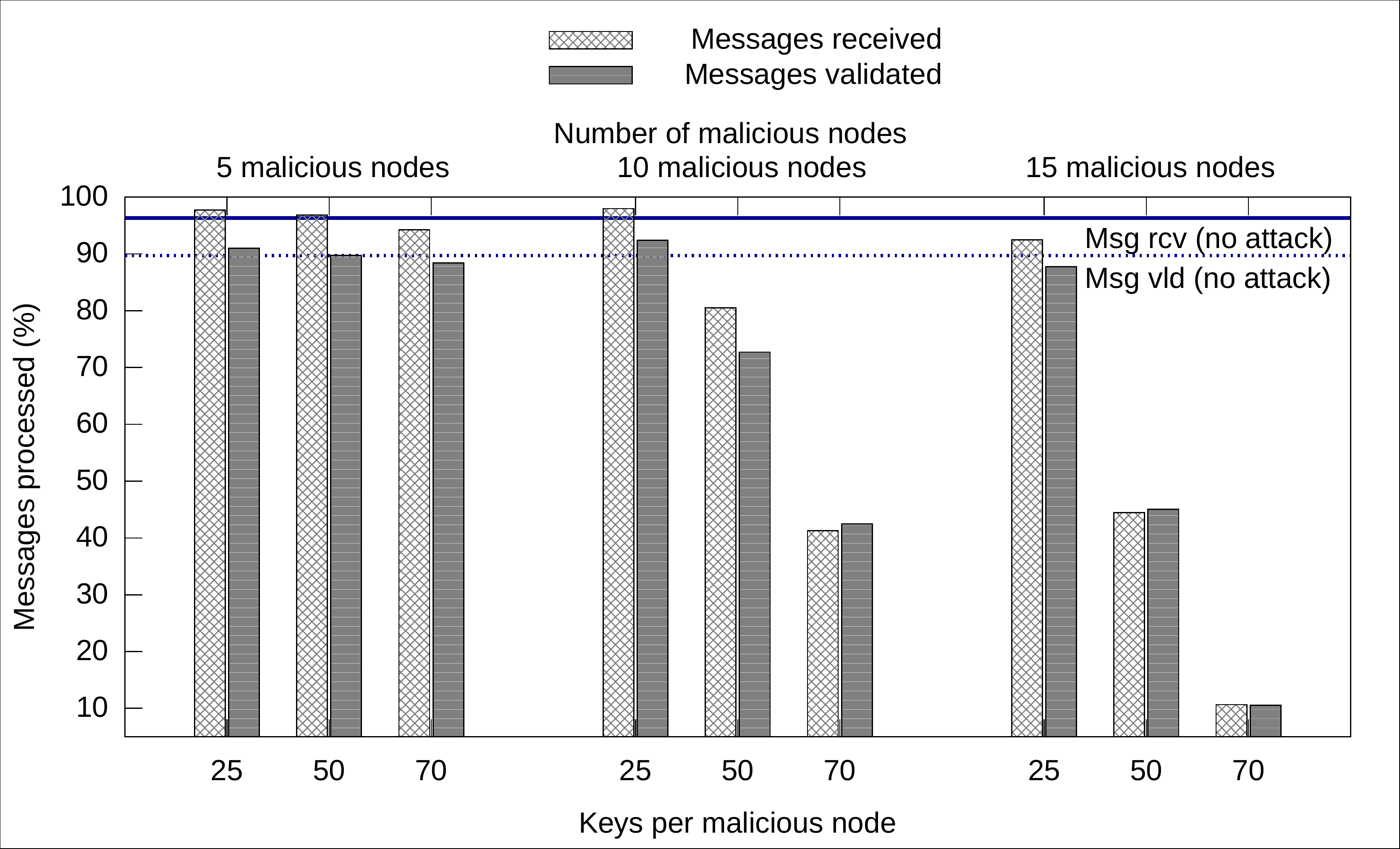}
	\caption{\ankit{Effect of the number of malicious nodes and keys per malicious nodes upon percentage of ``legitimate'' messages received and validated}}
	\label{chart:percentage}
\end{figure}

\ankit{The next important criteria to evaluate the impact of an attack on Algorand is the percentage of ``legitimate'' messages received and ``legitimate'' messages validated in due time. \figurename~\ref{chart:percentage} shows the effect of our attack upon ``legitimate'' messages received and validated with respect to the number of malicious nodes and keys per malicious nodes. Also here, with an attack payload of just 500 blocks (10 malicious nodes with 50 malicious keys), the percentage of ``legitimate'' messages received/validated starts to degrade considerably, which eventually lead to step failures.} 
\par
It is important to mention that an adversary can tune our attack on the number of malicious nodes/keys to prevent the detection while still creating significant disturbance in the network. Moreover, the size of the blocks is another important parameter to tune the attack. \rrmichele{
The results presented in the Figures~\ref{chart:time}~and~\ref{chart:percentage} were obtained using 1~MB of attack payload. Tables~\ref{table:block_size_timing}~and~\ref{table:block_size_messages} show the impact of attack payload's size upon the average round time and the percentage of ``legitimate'' messages received as well as validated in due time, respectively. 
Our results show that the size of the block does not significantly affect the performance of honest nodes in the absence of the attack. However, increasing size of the attack payload severely affect the performance of honest nodes in term of both the average round time and messages processed. It happens because the block's size directly affects the download time of a block, which is further worsened by the number of malicious nodes and the keys per malicious nodes involved in the attack.
}

\begin{table}[!htbp]
\centering
\caption{\rrankit{Effect of the attack payload's size upon average round time}}
\label{table:block_size_timing}
\resizebox{.95\columnwidth}{!}
	{
    \begin{tabular}{cc|c|c|c|c|}
\hline
\multicolumn{1}{|c|}{\textbf{\begin{tabular}[c]{@{}c@{}}Block size\\ (in MB)\end{tabular}}} & \textbf{\begin{tabular}[c]{@{}c@{}}Keys/malicious\\ node\end{tabular}} & \textbf{\begin{tabular}[c]{@{}c@{}}5 malicious\\  nodes\end{tabular}} & \textbf{\begin{tabular}[c]{@{}c@{}}10 malicious\\  nodes\end{tabular}} & \textbf{\begin{tabular}[c]{@{}c@{}}15 malicious\\  nodes\end{tabular}} & \textbf{\begin{tabular}[c]{@{}c@{}}No\\ attack\end{tabular}} \\ \hline
\multicolumn{1}{|c|}{\multirow{3}{*}{0.5}} & 25 & 178.65 & 180.55 & 182.36 & \multirow{3}{*}{181.42} \\ \cline{2-5}
\multicolumn{1}{|c|}{} & 50 & 180.73 & 181.36 & 181.72 &  \\ \cline{2-5}
\multicolumn{1}{|c|}{} & 70 & 178.98 & 180.45 & 237.27 &  \\ \hline
\multicolumn{1}{|c|}{\multirow{3}{*}{1.0}} & 25 & 182.21 & 182.87 & 183.24 & \multirow{3}{*}{181.16} \\ \cline{2-5}
\multicolumn{1}{|c|}{} & 50 & 182.49 & 232.15 & 391.16 &  \\ \cline{2-5}
\multicolumn{1}{|c|}{} & 70 & 181.69 & 391.12 & 391.45 &  \\ \hline
\multicolumn{1}{|c|}{\multirow{3}{*}{1.5}} & 25 & 180.16 & 180.97 & 241.52 & \multirow{3}{*}{181.54} \\ \cline{2-5}
\multicolumn{1}{|c|}{} & 50 & 181.22 & 391.58 & 391.66 &  \\ \cline{2-5}
\multicolumn{1}{|c|}{} & 70 & 235.76 & 391.28 & 391.54 &  \\ \hline
\multicolumn{1}{|c|}{\multirow{3}{*}{2.0}} & 25 & 178.97 & 226.63 & 391.31 & \multirow{3}{*}{182.31} \\ \cline{2-5}
\multicolumn{1}{|c|}{} & 50 & 226.59 & 390.88 & 391.57 &  \\ \cline{2-5}
\multicolumn{1}{|c|}{} & 70 & 391.39 & 391.83 & 391.93 &  \\ \hline
\multicolumn{1}{l}{} & \multicolumn{1}{l|}{\textbf{}} & \textbf{Time (s)} & \textbf{Time (s)} & \textbf{Time (s)} & \textbf{Time (s)} \\ \cline{3-6} 
\end{tabular}
    }
\end{table}

\begin{table*}[!htbp]
\centering
\caption{\rrankit{Effect of the attack payload's size upon the percentage of ``legitimate'' messages received and validated}}
\label{table:block_size_messages}
\resizebox{1.95\columnwidth}{!}
{  
    \begin{tabular}{cc|c|c|c|c|c|c|c|c|}
    \hline
    \multicolumn{1}{|c|}{\textbf{Block size (in MB)}} & \textbf{Keys/malicious node} & \multicolumn{2}{c|}{\textbf{5 malicious nodes}} & \multicolumn{2}{c|}{\textbf{10 malicious nodes}} & \multicolumn{2}{c|}{\textbf{15 malicious nodes}} & \multicolumn{2}{c|}{\textbf{No attack}} \\ \hline
    \multicolumn{1}{|c|}{\multirow{3}{*}{0.5}} & 25 & 96.98 & 91.61 & 97.26 & 92.49 & 97.08 & 90.16 & \multirow{3}{*}{97.51} & \multirow{3}{*}{90.53} \\ \cline{2-8}
    \multicolumn{1}{|c|}{} & 50 & 97.37 & 92.87 & 96.07 & 90.90 & 94.96 & 90.43 &  &  \\ \cline{2-8}
    \multicolumn{1}{|c|}{} & 70 & 97.85 & 91.25 & 96.04 & 90.21 & 75.03 & 64.99 &  &  \\ \hline
    \multicolumn{1}{|c|}{\multirow{3}{*}{1.0}} & 25 & 97.70 & 91.02 & 97.98 & 92.41 & 92.46 & 87.77 & \multirow{3}{*}{96.32} & \multirow{3}{*}{89.7} \\ \cline{2-8}
    \multicolumn{1}{|c|}{} & 50 & 96.81 & 89.75 & 80.52 & 72.72 & 45.47 & 44.06 &  &  \\ \cline{2-8}
    \multicolumn{1}{|c|}{} & 70 & 94.24 & 88.41 & 42.31 & 41.46 & 10.66 & 10.55 &  &  \\ \hline
    \multicolumn{1}{|c|}{\multirow{3}{*}{1.5}} & 25 & 99.32 & 94.85 & 93.62 & 89.97 & 65.62 & 63.32 & \multirow{3}{*}{97.43} & \multirow{3}{*}{90.77} \\ \cline{2-8}
    \multicolumn{1}{|c|}{} & 50 & 93.25 & 86.65 & 44.63 & 43.10 & 8.75 & 8.02 &  &  \\ \cline{2-8}
    \multicolumn{1}{|c|}{} & 70 & 73.57 & 67.28 & 25.93 & 24.32 & 6.79 & 5.82 &  &  \\ \hline
    \multicolumn{1}{|c|}{\multirow{3}{*}{2.0}} & 25 & 97.83 & 92.01 & 80.03 & 73.21 & 42.96 & 41.59 & \multirow{3}{*}{96.80} & \multirow{3}{*}{90.09} \\ \cline{2-8}
    \multicolumn{1}{|c|}{} & 50 & 79.99 & 75.12 & 26.66 & 25.21 & 1.05 & 1.00 &  &  \\ \cline{2-8}
    \multicolumn{1}{|c|}{} & 70 & 55.37 & 49.71 & 4.96 & 4.83 & 0.95 & 0.91 &  &  \\ \hline
     &  & \textbf{\begin{tabular}[c]{@{}c@{}}Messages\\ received (\%)\end{tabular}} & \textbf{\begin{tabular}[c]{@{}c@{}}Messages\\ validated (\%)\end{tabular}} & \textbf{\begin{tabular}[c]{@{}c@{}}Messages\\ received (\%)\end{tabular}} & \textbf{\begin{tabular}[c]{@{}c@{}}Messages\\ validated (\%)\end{tabular}} & \textbf{\begin{tabular}[c]{@{}c@{}}Messages\\ received (\%)\end{tabular}} & \textbf{\begin{tabular}[c]{@{}c@{}}Messages\\ validated (\%)\end{tabular}} & \textbf{\begin{tabular}[c]{@{}c@{}}Messages\\ received (\%)\end{tabular}} & \textbf{\begin{tabular}[c]{@{}c@{}}Messages\\ validated (\%)\end{tabular}} \\ \cline{3-10} 
    \end{tabular}
}
\end{table*}

\section{Feasibility of the attack}
\label{section:analysis}
\ankit{The attack costs practically nothing to an adversary. However, the adversary has to overcome one minor challenge, \ie to connect at least one malicious node with the honest target. This malicious node can then flood the target with an arbitrarily large number of block proposals - one for each public key controlled by the adversary. However, the recipient can monitor and label connections from which he receives a suspiciously large number of messages for a round. Then, it is likely that the honest nodes would simply drop (or, at least temporarily stop listening) from such a connection. Hence, the success of our attack depends on how many malicious connections an attacker is able to establish with the target node.
}
\par
The other important aspects of our attack are: (1)~since TCP connections are required to be established, IP spoofing is not an option\footnote{Establishing TCP connections via IP spoofing is problematic as one cannot complete the TCP three-way handshake protocol; because the responses (\ie the SYN-ACK packets) from the target nodes are not addressed to the real location of a malicious node. If we assume such possibility, then the adversary has control over at least one router on the path between the fake address and the target, or the target node's network has allowed source-routing~\cite{sunshine1977source} of the messages, or the Initial Sequence Number (ISN) in the handshake protocol is vulnerable to the prediction attack. All of these assumptions are quite strong hypothesis.} and real addresses should be used; 
(2)~the target's \textit{max-connections} parameter defines the maximum number of connections allowed towards it. We believe that the former issue can be solved by using botnet or relying on hidden services, such as Tor, to protect the address identity inside the network. However, the \textit{max-connections} parameter is out of the adversary's control. Hence, \michele{it is considered as an attack variable in our attack scenario. Consequently, establishing and maintaining a high number of connections with the target nodes is somehow a bit more challenging. But, the goal is achievable due to the current design of the protocol, where peers are not weighted on their stake.}

\section{Conclusion and future works}
\label{section:conclusion}
\ankit{The Algorand protocol was proposed to overcome the limitations of conventional blockchain technologies. The protocol has great potential and is in under active development.} In this paper, we present a particular security flaw of Algorand protocol that exploits its process of validating messages. In particular, we evaluated a possible DDoS-like flooding scenario \ankit{under practical assumptions}, where honest nodes suffer significant delays in the execution of protocol. Furthermore, we also discuss the major factors that make this attack scenario more challenging for the honest nodes. Finally, we also present possible solutions to prevent such an attack. The rigorous formalization of these countermeasures is kept as the future work.

\bibliographystyle{IEEEtran}
\bibliography{bib_full}

\begin{thebibliography}{10}
\providecommand{\url}[1]{#1}
\csname url@samestyle\endcsname
\providecommand{\newblock}{\relax}
\providecommand{\bibinfo}[2]{#2}
\providecommand{\BIBentrySTDinterwordspacing}{\spaceskip=0pt\relax}
\providecommand{\BIBentryALTinterwordstretchfactor}{4}
\providecommand{\BIBentryALTinterwordspacing}{\spaceskip=\fontdimen2\font plus
\BIBentryALTinterwordstretchfactor\fontdimen3\font minus
  \fontdimen4\font\relax}
\providecommand{\BIBforeignlanguage}[2]{{%
\expandafter\ifx\csname l@#1\endcsname\relax
\typeout{** WARNING: IEEEtran.bst: No hyphenation pattern has been}%
\typeout{** loaded for the language `#1'. Using the pattern for}%
\typeout{** the default language instead.}%
\else
\language=\csname l@#1\endcsname
\fi
#2}}
\providecommand{\BIBdecl}{\relax}
\BIBdecl

\bibitem{nakamoto2008bitcoin}
S.~Nakamoto, ``{Bitcoin: A Peer-to-Peer Electronic Cash System},'' [online]
  \url{https://bitcoin.org/bitcoin.pdf}, 2008.

\bibitem{conti2018economic}
M.~Conti, A.~Gangwal, and S.~Ruj, ``{On the Economic Significance of Ransomware
  Campaigns: A Bitcoin Transactions Perspective},'' \emph{Elsevier Computers \&
  Security}, vol.~79, pp. 162--189, 2018.

\bibitem{haber1990time}
S.~Haber and W.~S. Stornetta, ``{How to Time-stamp a Digital Document},'' in
  \emph{Conference on the Theory and Application of Cryptography (CRYPTO)},
  1990, pp. 437--455.

\bibitem{szabo2005bitgold}
N.~Szabo, ``{Bit Gold},'' [online]
  \url{https://unenumerated.blogspot.com/2005/12/bit-gold.html}, 2005.

\bibitem{Wei1998money}
W.~Dai, ``{b-money},'' [online] \url{http://www.weidai.com/bmoney.txt}, 1998.

\bibitem{finney2004rpow}
H.~Finney, ``{RPOW - Reusable Proof Of Work},'' [online]
  \url{http://cryptome.org/rpow.htm}, 2004.

\bibitem{blockchain_app}
``{Banking Is Only The Beginning: 42 Big Industries Blockchain Could
  Transform},'' [online]
  \url{https://www.cbinsights.com/research/industries-disrupted-blockchain/},
  2018.

\bibitem{croman2016scaling}
K.~Croman, C.~Decker, I.~Eyal, A.~E. Gencer, A.~Juels, A.~Kosba, A.~Miller,
  P.~Saxena, E.~Shi, E.~G. Sirer \emph{et~al.}, ``{On Scaling Decentralized
  Blockchains},'' in \emph{20th International Conference on Financial
  Cryptography and Data Security}, 2016, pp. 106--125.

\bibitem{rosenfeld2011analysis}
M.~Rosenfeld, ``{Analysis of Bitcoin Pooled Mining Reward Systems},''
  \emph{arXiv preprint arXiv:1112.4980}, 2011.

\bibitem{eyal2015miner}
I.~Eyal, ``{The Miner's Dilemma},'' in \emph{36th IEEE Symposium on Security
  and Privacy (SP)}, 2015, pp. 89--103.

\bibitem{luu2015power}
L.~Luu, R.~Saha, I.~Parameshwaran, P.~Saxena, and A.~Hobor, ``{On Power
  Splitting Games in Distributed Computation: The Case of Bitcoin Pooled
  Mining},'' in \emph{28th IEEE Computer Security Foundations Symposium (CSF)},
  2015, pp. 397--411.

\bibitem{malone2014energy}
K.~J. O'Dwyer and D.~Malone, ``{Bitcoin Mining and its Energy Footprint},''
  2014.

\bibitem{percival2009stronger}
C.~Percival, ``{Stronger Key Derivation via Sequential Memory-hard
  Functions},'' [online] \url{https://www.tarsnap.com/scrypt/scrypt.pdf}, pp.
  1--16, 2009.

\bibitem{king2013primecoin}
S.~King, ``{Primecoin: Cryptocurrency with Prime Number Proof-of-Work},''
  [online] \url{http://primecoin.io/bin/primecoin-paper.pdf}, 2013.

\bibitem{wiki_pob}
I.~Stewart, ``{Proof of Burn},'' [online]
  \url{https://en.bitcoin.it/wiki/Proof_of_burn}.

\bibitem{slimcoin}
P4Titan, ``{Slimcoin. A Peer-to-Peer Crypto-Currency with Proof-of-Burn},''
  [online]
  \url{https://github.com/slimcoin-project/slimcoin-project.github.io/raw/master/whitepaperSLM.pdf}.

\bibitem{intel}
``{Hyperledger Sawtooth documentation},'' [online]
  \url{https://intelledger.github.io/}, 2016.

\bibitem{king2012ppcoin}
S.~King and S.~Nadal, ``{PPcoin: Peer-to-Peer Crypto-currency with
  Proof-of-Stake},'' [online]
  \url{https://pdfs.semanticscholar.org/0db3/8d32069f3341d34c35085dc009a85ba13c13.pdf},
  2012.

\bibitem{cloakcoin2014}
``{Enigma. A Private, Secure and Untraceable Transactions System for
  Cloackcoin},'' [online]
  \url{https://www.cloakcoin.com/user/themes/g5_cloak/resources/CloakCoin_Whitepaper_v2.1.pdf},
  2014.

\bibitem{ren2014proof}
L.~Ren, ``{Proof of Stake Velocity: Building the Social Currency of the Digital
  Age},'' [online] \url{https://www.reddcoin.com/papers/PoSV.pdf}, 2014.

\bibitem{pikepost}
D.~Pike, P.~Nosker, D.~Boehm, D.~Grisham, S.~Woods, and J.~Marston, ``{PoST
  White Paper},'' [online]
  \url{https://cdn.vericonomy.com/documents/VeriCoin-Proof-of-Stake-Time-Whitepaper.pdf}.

\bibitem{nxt}
``{Whitepaper: Nxt},'' [online]
  \url{https://nxtwiki.org/wiki/Whitepaper:Nxt#Nxt.E2.80.99s_Proof_of_Stake_Model},
  2014.

\bibitem{vasin2014blackcoin}
P.~Vasin, ``{Blackcoin's Proof-of-Stake Protocol V2},'' [online]
  \url{https://blackcoin.org/blackcoin-pos-protocol-v2-whitepaper.pdf}, 2014.

\bibitem{li2017securing}
W.~Li, S.~Andreina, J.-M. Bohli, and G.~Karame, ``{Securing Proof-of-Stake
  Blockchain Protocols},'' in \emph{Data Privacy Management - Cryptocurrencies
  and Blockchain Technology (DPM - CBT 2017)}, 2017, pp. 297--315.

\bibitem{kiayias2017ouroboros}
A.~Kiayias, A.~Russell, B.~David, and R.~Oliynykov, ``{Ouroboros: A Provably
  Secure Proof-of-Stake Blockchain Protocol},'' in \emph{Annual International
  Cryptology Conference (CRYPTO)}.\hskip 1em plus 0.5em minus 0.4em\relax
  Springer, 2017, pp. 357--388.

\bibitem{bentov2016snow}
I.~Bentov, R.~Pass, and E.~Shi, ``{Snow White: Provably Secure
  Proofs-of-Stake},'' \emph{IACR Cryptology ePrint Archive}, vol. 2016, p. 919,
  2016.

\bibitem{buterin2017casper}
V.~Buterin and V.~Griffith, ``{Casper the Friendly Finality Gadget},''
  \emph{arXiv preprint arXiv:1710.09437}, 2017.

\bibitem{buterin2014slasher}
V.~Buterin, ``{Slasher: A Punitive Proof-of-Stake Algorithm},'' [online]
  \url{https://ethereum.github.io/blog/2014/01/15/slasher-a-punitive-proof-of-stake-algorithm/},
  2015.

\bibitem{sompolinsky2016spectre}
Y.~Sompolinsky, Y.~Lewenberg, and A.~Zohar, ``{SPECTRE: A Fast and Scalable
  Cryptocurrency Protocol},'' \emph{IACR Cryptology ePrint Archive}, p. 1159,
  2016.

\bibitem{popov2016iota}
S.~Popov, ``{IOTA: The tangle},'' 2016.

\bibitem{micali2016algorand_tec}
J.~Chen and S.~Micali, ``Algorand,'' \emph{arXiv preprint arXiv:1607.01341},
  2016.

\bibitem{micali2017algorand_pap}
Y.~Gilad, R.~Hemo, S.~Micali, G.~Vlachos, and N.~Zeldovich, ``{Algorand:
  Scaling Byzantine Agreements for Cryptocurrencies},'' in \emph{26th ACM
  Symposium on Operating Systems Principles}, 2017, pp. 51--68.

\bibitem{micali1999verifiable}
S.~Micali, M.~Rabin, and S.~Vadhan, ``{Verifiable Random Functions},'' in
  \emph{40th IEEE Annual Symposium on Foundations of Computer Science}, 1999,
  pp. 120--130.

\bibitem{openstack}
``{OpenStack},'' [online] \url{www.openstack.org}, 2018.

\bibitem{sunshine1977source}
C.~A. Sunshine, ``{Source Routing in Computer Networks},'' \emph{ACM SIGCOMM
  Computer Communication Review}, vol.~7, no.~1, pp. 29--33, 1977.

\end{thebibliography}

%


\begin{IEEEbiography}[{\includegraphics[width=1in,height=1.25in,clip
			]{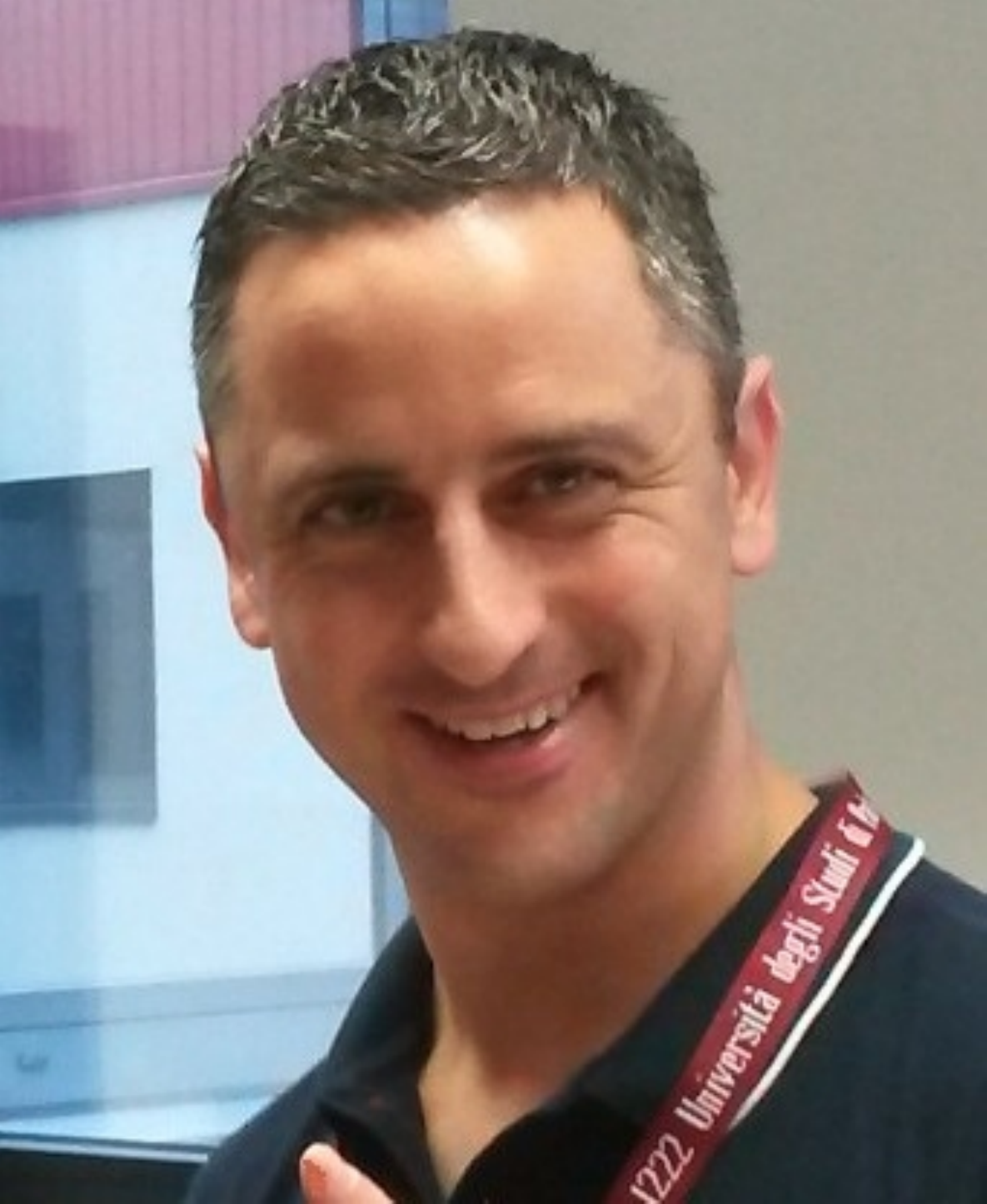}}]{Mauro Conti} is Full Professor at the University of Padua, Italy, and Affiliate Professor at the University of Washington, Seattle, USA. He obtained his Ph.D. from Sapienza University of Rome, Italy, in 2009. After his Ph.D., he was a Post-Doc Researcher at Vrije Universiteit Amsterdam, The Netherlands. In 2011 he joined as Assistant Professor the University of Padua, where he became Associate Professor in 2015, and Full Professor in 2018. He has been Visiting Researcher at GMU (2008, 2016), UCLA (2010), UCI (2012, 2013, 2014, 2017), TU Darmstadt (2013), UF (2015), and FIU (2015, 2016, 2018). He has been awarded with a Marie Curie Fellowship (2012) by the European Commission, and with a Fellowship by the German DAAD (2013). His research is also funded by companies, including Cisco, Intel, and Huawei. His main research interest is in the area of security and privacy. In this area, he published more than 250 papers in topmost international peer-reviewed journals and conference. He is Area Editor-in-Chief for IEEE Communications Surveys \& Tutorials, and Associate Editor for several journals, including IEEE Communications Surveys \& Tutorials, IEEE Transactions on Information Forensics and Security, IEEE Transactions on Dependable and Secure Computing, and IEEE Transactions on Network and Service Management. He was Program Chair for TRUST 2015, ICISS 2016, WiSec 2017, and General Chair for SecureComm 2012 and ACM SACMAT 2013. He is Senior Member of the IEEE.
	\end{IEEEbiography}
	
	\begin{IEEEbiography}[{\includegraphics[width=1in,height=1.25in,clip]{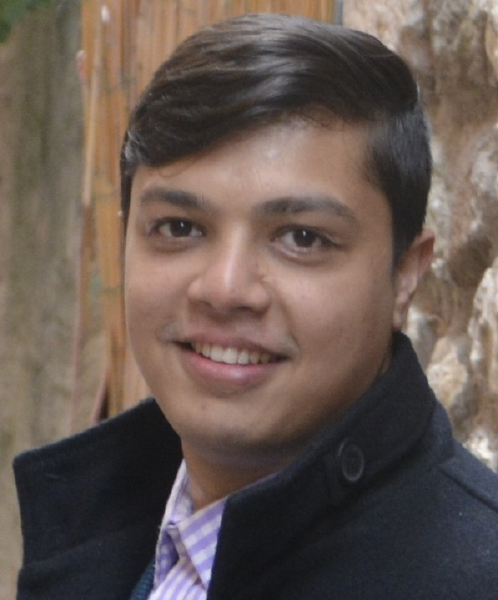}}]{Ankit Gangwal} received the B.Tech. degree in Information Technology from RTU Kota, India, in 2011 and the M.Tech. degree in Computer Engineering from Malaviya National Institute of Technology Jaipur, India, in 2016. Currently, he is a Ph.D. student in the Department of Mathematics, University of Padua, Italy with a fellowship for international students funded by Fondazione Cassa di Risparmio di Padova e Rovigo (CARIPARO). His~current research interest is in the area of security and privacy of the blockchain technology and novel network architectures, in particular, software-defined network.
	\end{IEEEbiography}

	\begin{IEEEbiography}[{\includegraphics[width=1in,height=1.25in,clip]{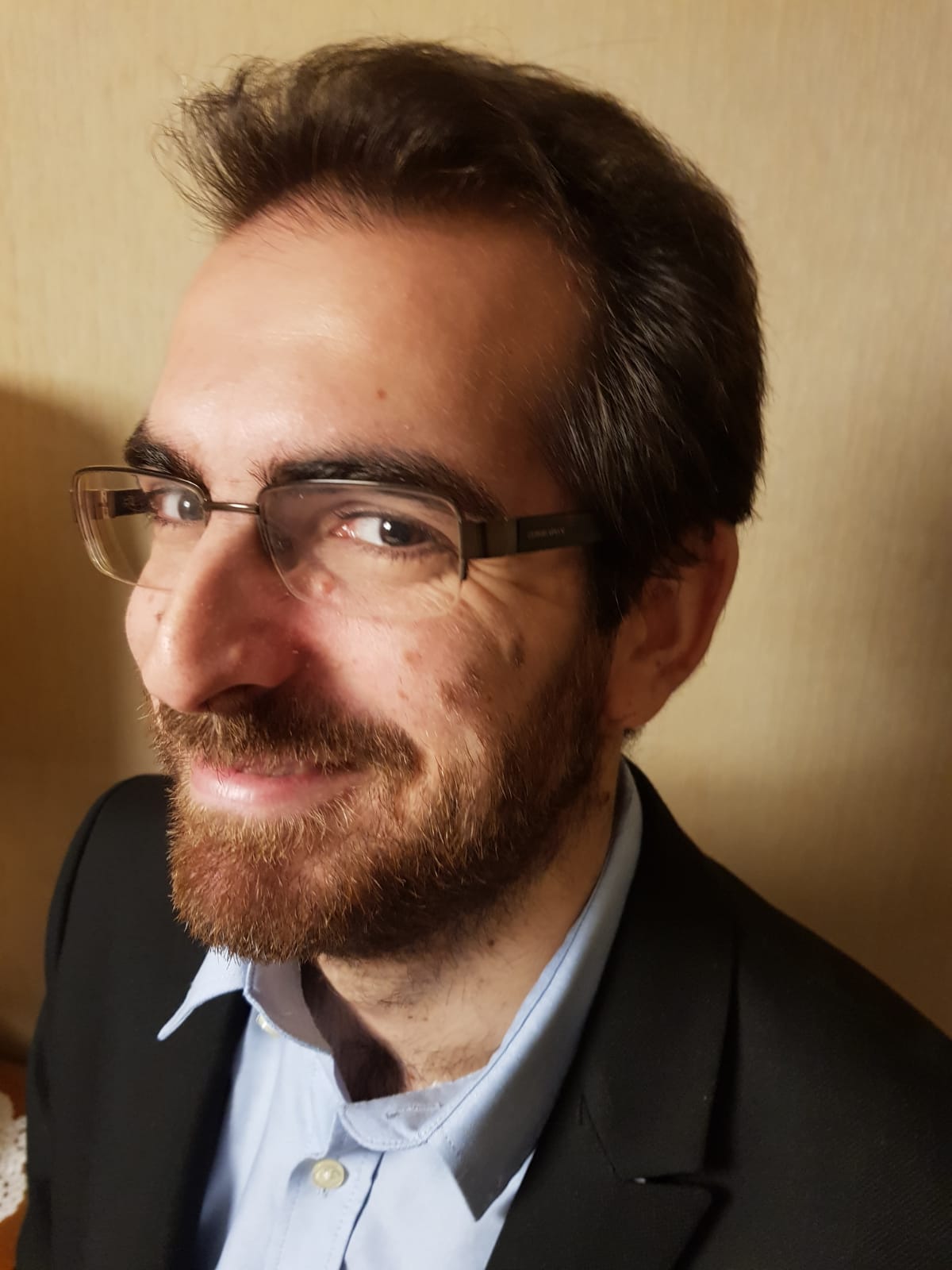}}]{Michele Todero} has received his M.Sc. (2018) and B.Sc. (2015) in Computer Engineering from the Department of Information Engineering of University of Padua, Italy. His current research focus on security and analysis of innovative blockchain consensus algorithms.
	\end{IEEEbiography}
	
	\balance
\end{document}